# New Stacking Variations of the CE-type Structure in the Metal-Ordered Manganite $YBaMn_2O_6$

[1]H. Kageyama, [1]T. Nakajima, [1]M. Ichihara, [1]Y. Ueda, and [2]H. Yoshizawa, [3]K. Ohoyama

[1] *Material Design and Characterization Laboratory, Institute for Solid State Physics, University of Tokyo, 5-1-5 Kashiwanoha, Kashiwa, Chiba 277-8581*

[2] *Neutron Scattering Laboratory, Institute for Solid State Physics, The University of Tokyo, 106-1 Shirakata, Tokai, Ibaraki 319-1106*

[3] *Institute for Materials Research, Tohoku University, 2-1-1 Katahira, Aoba-ku, Sendai 980-857*

Abstract

An *A*-site ordered manganese perovskite $YBaMn_2O_6$ undergoes unusual and multiple phase transitions; a structural transition at $T_{c1}$ = 520 K, a metal-insulator transition at $T_{c2}$ = 480 K and an antiferromagnetic long-range order at $T_{c3}$ = 195 K. We performed transmission electron microscope (TEM) and powder neutron diffraction experiments, particularly focusing on how charge, orbital and spin degrees of freedom freeze at low temperatures. For the paramagnetic insulating phase ($T_{c3}<T<T_{c2}$), the so-called CE type of charge and orbital ordered state was observed within the monoclinic *a-b* plane, which is most commonly observed for the ordinary solid solution of $A^{3+}_{1-x}A'^{2+}_{x}MnO_3$ around $x$ = 0.5. However, TEM revealed a 4-fold periodicity along the *c* axis, suggesting a new stacking pattern, where planes of the CE type are built up according to the sequence [    ...]. Interestingly, when the system enters into the antiferromagnetic state below $T_{c3}$, this stacking pattern changes into [    ...] or [    ...], suggesting close interplay between spins and orbitals. These features are  discussed in terms of inherent structural alternation, i.e., the Y/Ba order along the *c* axis.

Perovskite ($ABO_3$) and its related compounds display interesting electric, magnetic and optical properties. Most well-known examples are the high-$T_c$ superconductivity in cuprates ($B$=Cu) [1] and the colossal magnetoresistance in manganites ($B$=Mn) [2]. Emergence of various phenomena of interests in the perovskite systems is partly due to the versatility of the $A$ site to form a wide range of solid solution between cations with different valences and ionic sizes. This advantage provides us the opportunity to systematically tune doping level $x$ as well as the Goldschmidt tolerance factor $f$, as demonstrated by $La_{2-x}Sr_xCuO_4$ and $La_{1-x}Ca_xMnO_3$ [3, 4].

It should be stressed, however, that such partial substitution can always be a cause of structural disorder. In $A_{1-x}A'_xMnO_3$, the shape, size and tilting angle of each $MnO_6$ octahedron would depend largely on nearby $A/A'$ ions, so that crystallographic parameters refined usually using X-ray or neutron diffraction data describe nothing but specially averaged ones. A problem is that randomness of the underlying lattice should affect the physical properties more or less. The effect would become more serious when one deals with microscopic phenomena. It is possible that the structural disorder and spacial heterogeneity trigger or accelerate the nucleation of microscopic electronic phase segregations recently observed in $La_{2-x}Sr_xCuO_4$ [5] and $La_{0.5}Sr_{0.5}MnO_3$ [6].

Fortunately, one can tailor a 'clean' system only when $x$ in $A^{3+}_{1-x}A'^{2+}_xBO_3$ is 1/2. Taking an example of $Y_{0.5}Ba_{0.5}CoO_{3-y}$ [7, 8], the $A$-site metal order takes place in a way that the YO and BaO layers are built up alternatively. Therefore, the composition should be better denoted as $YBaCo_2O_{6-y}$. Recently, we have succeeded to synthesize its manganese analogue, $YBaMn_2O_6$ ($A$ = Y, $A'$ = Ba, $B$ = Mn) [9, 10], whose structural and physical properties are considerably different from the conventional disordered $A_{0.5}A'_{0.5}MnO_3$ system with the $A/A'$ ions being randomly distributed [11, 12]. $YBaMn_2O_6$ undergoes -with descending temperature $T$- unusual phase transitions; a 1st order structural phase transition at $T_{c1}$ = 520 K from triclinic $P1$ to monoclinic $P2$, a metal-insulator (MI) transition at $T_{c2}$ = 480 K and an antiferromagnetic ordering at $T_{c3}$ = 195 K. A peculiar feature is the separation of structural and MI transitions (i.e., $T_{c2}$ $T_{c1}$), strikingly in contrast to the $A_{0.5}A'_{0.5}MnO_3$ system, where these transitions occurs simultaneously (i.e., $T_{c1}$ = $T_{c2}$). Moreover, the value of $T_{c2}$ is, to the authors knowledge, highest in the manganese perovskite family.

Structural considerations [9, 10] naively tell us that these unusual properties in $YBaMn_2O_6$ are attributable to the absence of lattice disorder, the two-dimensionality due to the stacking sequence of $-YO-MnO_2-BaO-MnO_2-$, and rather distorted $MnO_6$ octahedron from ideal cubic one. However, for the fully understanding of the role of the structure (especially, order/disorder), it is necessary to determine the ordered superstructures of charge, orbital and spin degrees of freedom at low temperatures. In this study, we performed transmission electron microscope (TEM) and neutron powder diffraction for the paramagnetic insulator (PI; $T_{c3} < T < T_{c2}$) and the antiferromagnetic insulator (AFI; $T < T_{c3}$). It is found that the obtained CO ordered pattern in the PI phase differs from those for the $A_{0.5}A'_{0.5}MnO_3$ systems. Interestingly, the antiferromagnetic long-range order redesigns the CO ordered pattern.

Polycrystalline sample of $YBaMn_2O_6$ was prepared by a conventional solid state reaction as reported in Ref. [9]. The superlattice originating from the CO order was probed by means of TEM using JEM2010 (JOEL) operating at 200 kV, equipped with a low-temperature sample holder, from the room temperature (RT) down to 24 K. The sample was finely ground under methanol then dispersed on Cu grids coated with holy-carbon support films. The neutron powder diffraction experiments, which allow us to determine the magnetic ordering, were conducted for 25 K $< T <$ 350 K on the Kinken powder diffractometer for high efficiency and high resolution measurements, HERMES, of Institute for Materials Research (IMR), Tohoku University, installed at the JRR-3M reactor in Japan Atomic Energy Research Institute (JAERI), Tokai. Neutrons with a wavelength of 1.8196 Å were obtained by the 331 reflection of the Ge monochromator and 12'-blank-sample-18' collimation.

Shown in Fig. 1 is the crystal structure of $YBaMn_2O_6$ at 350 K (the PI phase) determined from the Rietveld refinement of both powder X-ray and neutron diffraction patterns. The details of the structural analysis will be reported elsewhere [10]. The structure of the PI phase adopts a monoclinic space group $P2$ with a unit cell dimension of $2a_p \times 2b_p \times 2c_p$, where $a_p$, $b_p$ and $c_p$ denote the cell constants for the simple cubic perovskite. The lattice parameters are determined to be $a = 5.5193$ Å, $b = 5.5131$ Å, $c = 7.6135$ Å, and $= 90.295$ °.

There are two crystallographically inequivalent Mn sites, Mn(1) and Mn(2). There is a sizable difference of the volume of the $MnO_6$ octahedra (9.56 $Å^3$ for Mn(1) and 10.00 $Å^3$ for Mn(2)), suggesting a charge ordering that the $Mn^{3+}$ and $Mn^{4+}$ ions occupy the Mn(2) and Mn(1) sites, respectively. Each $MnO_6$ octahedron is heavily distorted (see Fig. 1) as a result of the A-site order, and the octahedral tilting system is written approximately as $a^-b^-c^-$.

Figure 2 (a) is the $[100]_p$-zone TEM image and corresponding electron diffraction pattern for the PI phase, which clearly reveals a modulation over a wide range of the specimen, originating from the successful alternation of the YO and BO layers along the c axis. In fact, the 7.5 Å-periodicity of the modulation is in good accordance with $c = 7.6135$ Å. The doubling of the c parameter can be confirmed by the electron diffraction pattern, which contains, in addition to the main reflection spots, relatively strong superlattice reflections with a 2-fold periodicity along $[001]_p$. Furthermore, we observed very weak reflections, the modulation (*q*) of which is commensurate and can be denoted as $q_1 = (0, 0, 1/4)_p$. The quadruplication along the c axis was invisible from the X-ray and neutron diffraction experiments, and its origin will be discussed later,

Direct observation of the CO states can be probed by the TEM investigations [13-15]. In the inset of Fig. 2 (b), we show the $[001]_p$-zone electron diffraction pattern taken at RT, exhibiting the satellite reflections of the CO ordering of a modulation following with the main reflection spots. This modulation appears along $[110]_p$ (or $[1-10]_p$) direction within the $a^*$-$b^*$ plane, and the structural modulation can be written as $q_2 = (1/4, 1/4, 0)_p$. The corresponding lattice image (Fig. 2 (b)) indicates the contrast of the $q_2$ superlattice running diagonally with respect to the $a_p$ and $b_p$ axes. In addition, one can see the contrast of the lattice constants $a_p$ and $b_p$ (~5.5 Å). The $q_2$ superlattice can be ascribed to the CO ordering of the so-called CE type in the a-b plane, namely, the arrangement of the $Mn^{3+}/Mn^{4+}$ ions into the $MnO_2$ square lattice accompanied by the alternative arrangement of rows of $d_{3x^2-r^2}$ and $d_{3y^2-r^2}$ orbitals running either along $[110]_p$ or $[1-10]_p$. The same pattern has been most frequently obtained for $A_{1-x}A'_xMnO_3$ when x is around 0.5 [13, 14].

Together with electron diffractions along various directions, we plotted thus obtained superspots on the reciprocal lattice in Fig. 3 (a). The most intriguing feature in the present study is the presence of the modulation vector $q_1 = (0, 0, 1/4)_p$. We interpret the origin of the $q_1$ modulated structure in terms of stacking sequence of the CE-type layers. By utilizing the interconversion of the $Mn^{3+}$- and $Mn^{4+}$- sublattices or/and the $d_{3x^2-r^2}$- and $d_{3y^2-r^2}$-sublattices, one can pile up the CE-type layers along the $c$ axis so as to satisfy the unit cell $2 \cdot 2a_p \times 2b_p \times 4c_p$. However, the crystal structure for the PI state (Fig. 1) excludes models that contain the operation of the interconversion between the $Mn^{3+}$- and $Mn^{4+}$- sublattices. Consequently, one can uniquely obtain a stacking sequence of [ ...] as depicted in Fig. 4 (a), although there are still two choices that the BaO layers are located between out-of-phase planes, i.e.,    and between in-phase planes, i.e.,   ( ).

This extended three-dimensional (3D) CE pattern is observed for the first time and forms a striking contrast to the half-doped solid solution, where an uniformly stacked pattern (Fig. 4 (b)), the original 3D CE pattern given by Wollan *et al.* [16] has always been observed [13, 14]. It is noted that the unusual stacking sequence [ ...] is also observed for the 2D system $NaV_2O_5$. In this vanadate, a number of *charge* stacking patterns appear with increasing pressure, which is successfully mapped onto the ANNNI model (devil's staircase) [17]. In the manganese perovskite systems, a most probable key in determining the 3D CO ordered structures should lie in the *A*-site order/disorder. In case of $A_{0.5}A'_{0.5}MnO_3$, the *orbital* interlayer interaction that governs stacking form is uniform, thus providing two possible superstructures shown in Figs. 4 (b) and (c), though the real system favors the former as mentioned above. On the other hand, such interlayer interactions should be alternated owing to the fact that the $MnO_2$ layers are sandwiched by the YO and BaO layers with considerably different ionic radii ($r_{Y^{3+}} = 1.18$ Å, $r_{Ba^{2+}} = 1.61$ Å), thus giving rise to the superlattice in Fig. 4 (a).

Finally, we would like to discuss the superstructure in the AFI state below $T_{c3}$. Figure 5 compares the neutron diffraction patterns taken at 250 K and 20 K. While all of the peaks for the PI phase are of nuclear origin and can be indexed with a monoclinic symmetry with the $2a_p \times 2b_p \times 2c_p$ unit cell, additional peaks of magnetic origin appear when $T$ is decreased below $T_{c3}$. It is revealed from indexing these magnetic peaks the spin-ordering pattern in the *a-b* plane is the same as that observed in $A_{0.5}A'_{0.5}MnO_3$ [18], but that along the $c$ axis has a 4-fold periodicity. A proposed spin structure is shown in Fig. 4 (d). Such quadruplication along the $c$ axis has not been observed for the $A_{0.5}A'_{0.5}MnO_3$ system. Again, it is quite natural to consider that the Y/Ba order along the $c$ axis quite gives rise to the alternation of the *magnetic* interlayer interactions and hence leads to such a periodicity.

The electron diffraction pattern at 24 K (Fig. 2 (c)) is interpreted as the superimposition of twin domains both having the $q_2$ modulation. Thus the CE type of the CO ordering is retained within the $a$-$b$ plane. Figure 2 (d) shows, interestingly enough, the disappearance of the modulation vector of $q_1$ at 24 K, although the system still has the 4-fold periodicity due to the spin ordering. The superspots at 24K obtained by TEM is summarized in Fig. 3 (b) and possible models of the CO ordering are depicted in Figs. 4 (b) and (c), where layers of the CE type are piled up according to [   ...] and [   ...], respectively. The change of the 3D superstructures must be a consequence of the competition between the CO ordering and the spin ordering. This rearrangement across $T_{c3}$ is partly supported by the nature of the 1st-order phase transition (relatively large hysteretic behavior in magnetic susceptibility and heat absorption/release by a differential scanning calorimetry (DSC) measurement) [9].

To summarize, we have investigated the superstructures of $YBaMn_2O_6$ in the charge and orbital ordered states by means of TEM. In addition to the superlattice reflection spots originating from the $A$-site metal order, those with modulation vectors $q_1 = (0, 0, 1/4)_p$ and $q_2 = (1/4, 1/4, 0)_p$ appeared in the PI phase. This modulation structure was interpreted taking into account the charge and orbital ordering of the CE type and its stacking arrangement. The stacking sequence according to [   ...] was first observed among the half-doped manganese perovskite compounds. In the AFI phase, we observed the disappearance of $q_1$, indicating the uniform [   ...] or alternative [   ...] stacking along the $c$ axis. Transformation of the charge/orbital ordering is mediated by the spin ordering with a quadruplicate periodicity along the same axis.

This work is partly supported by Grant-in-Aid for Scientific Research No. 40302640, No. 407 and No. 758 and for Creative Scientific Research (No. 13NP0201) from Ministry of Education, Science, Technology, Sports and Culture,

Japan.

Figure Captions

Fig. 1: A perspective view of the crystal structure of the monoclinic $YBaMn_2O_6$ at 350 K (the PI phase) looking down $[010]_p$, where the $Mn(1)O_6$ and $Mn(2)O_6$ octahedra are shown as light and dark polyhedra, respectively, while the $Ba^{2+}$ and $Y^{3+}$ ions are shown in dark and light balls.

Fig. 2: High-resolution TEM images and electron diffraction patterns of $YBaMn_2O_6$ at RT taken along (a) $[100]_p$-zone and (b) $[001]_p$-zone axes.

Electron diffraction patterns at 24 K taken along (c) $[001]_p$-zone and (d) $[100]_p$-zone axes.

Fig. 3: The reciprocal lattices with fundamental (closed) and superlattice (open) reflections, obtained from TEM at (a) RT and (b) 24 K.

Fig. 4: Schematic representation of charge and orbital ordering patterns (a)-(c) and of spin-ordering pattern (d) for $YBaMn_2O_6$, where for simplicity only $Mn^{3+}$ and $Mn^{4+}$ ions are drawn. (a): Proposed model for the PI state. (b) and (c): Proposed models for the AFI state. It is noted that the model (b) having uniform stacking is the original 3D CE structure and is obtained for the metal-disordered $A_{0.5}A'_{0.5}MnO_3$ system.

Fig. 5: Neutron diffraction profiles of $YBaMn_2O_6$ at 250 K and 20 K, where the peaks are indexed according to the $2a_p \times 2b_p \times 2c_p$ unit cell.

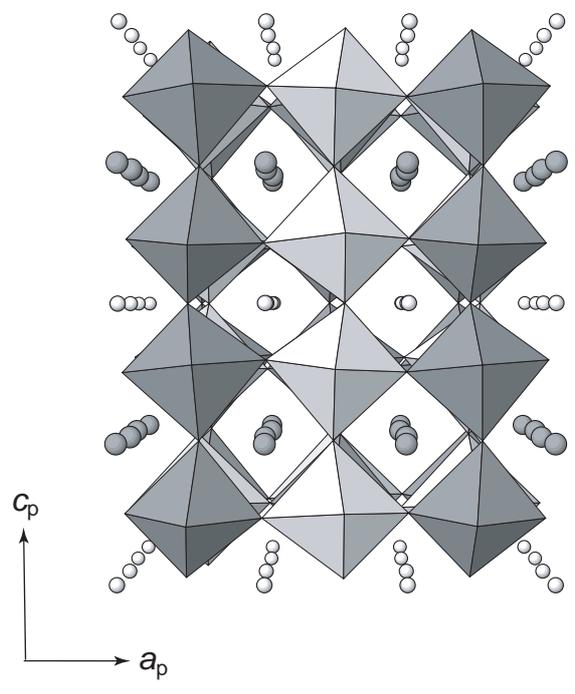

Fig.1 H. Kageyama *et al*.

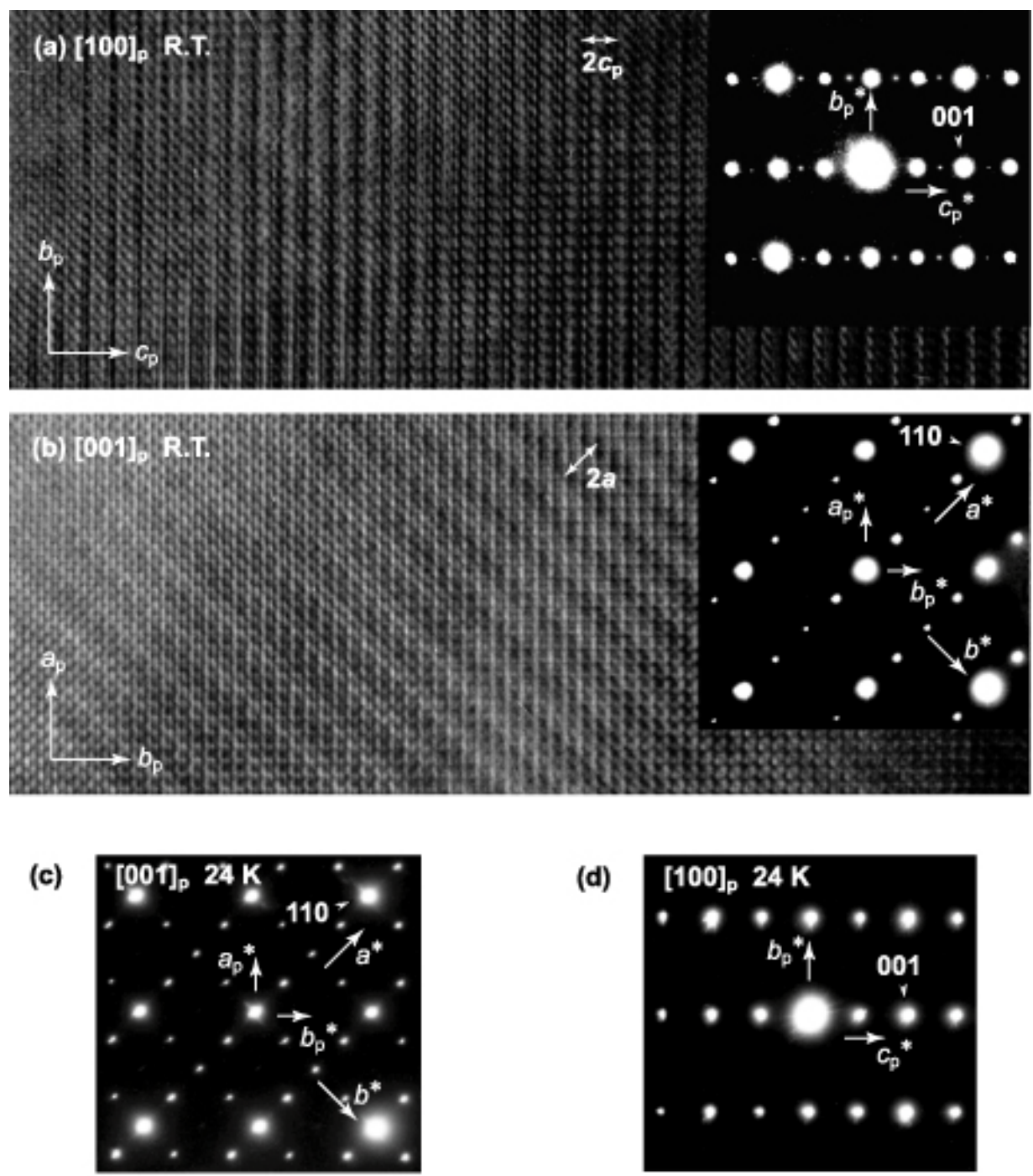

Fig.2 H. Kageyama et al.

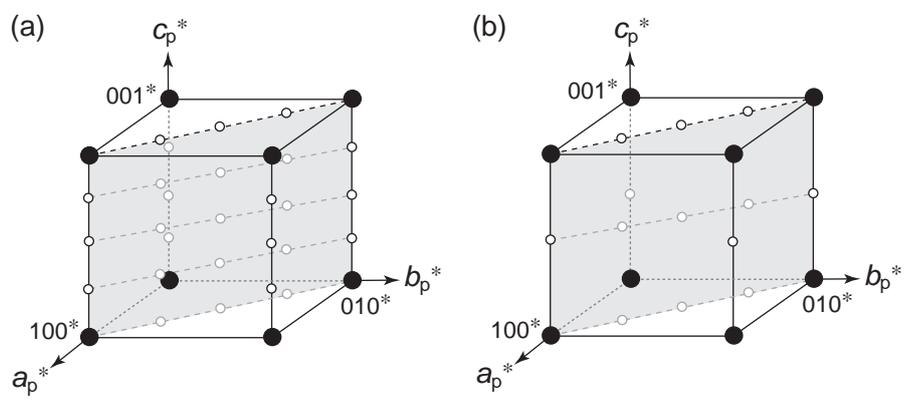

Fig.3 H. Kageyama *et al*.

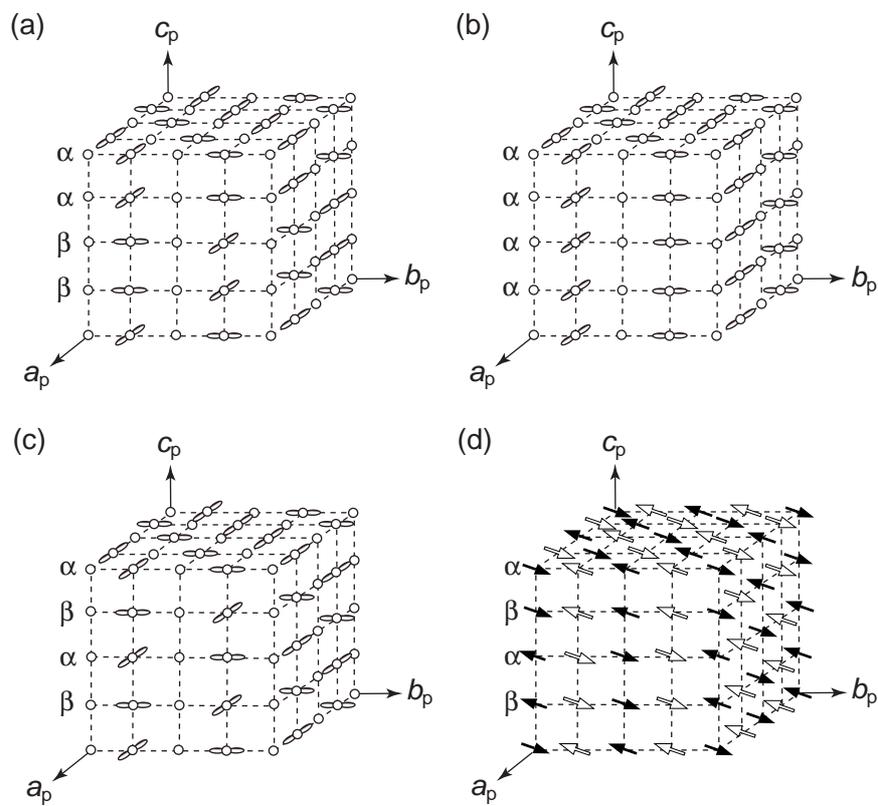

Fig. 4 H. Kageyama *et al.*

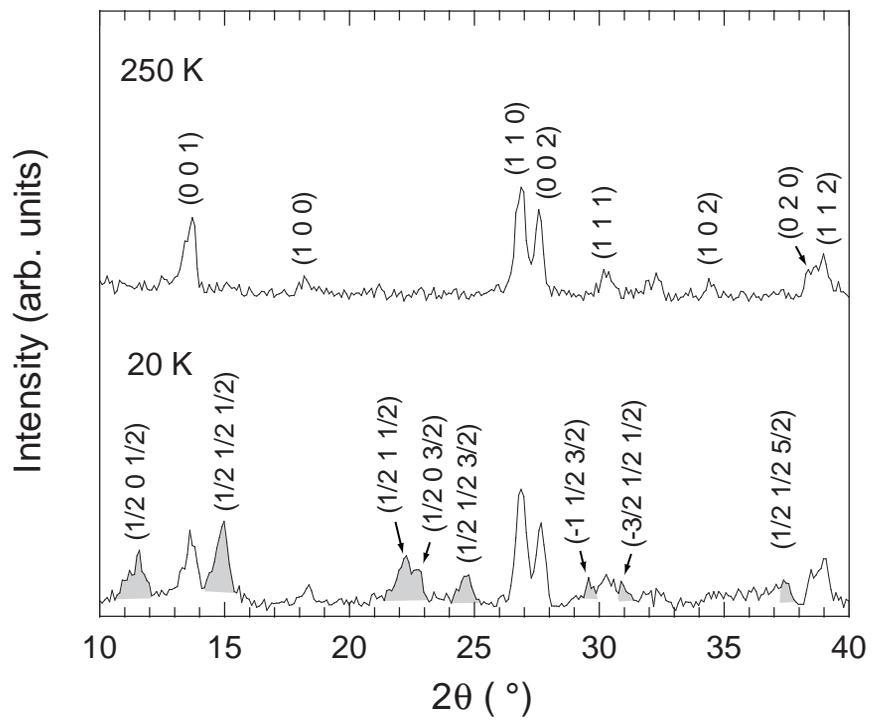

Fig.5  H. Kageyama  *et al.*